\documentclass[prb,twocolumn,amsmath,amssymb,showpacs]{revtex4}

\usepackage{graphicx}

\begin{document}

\newcommand{\etal}{{\it et al.}\/}

\newcommand{\gtwid}{\mathrel{\raise.3ex\hbox{$>$\kern-.75em\lower1ex\hbox{$\sim$}}}}

\newcommand{\ltwid}{\mathrel{\raise.3ex\hbox{$<$\kern-.75em\lower1ex\hbox{$\sim$}}}}

\title{Cold Attractive Spin Polarized Fermi Lattice Gases and the Doped 
Positive U Hubbard Model}

\author{Adriana Moreo$^{1}$ and D.~J.~Scalapino$^{2}$} 

\address{$^1$Department of Physics and Astronomy,University of Tennessee,
Knoxville, TN 37966-1200 and
\\Oak Ridge National Laboratory,Oak Ridge,
TN 37831-6032,USA, }
  
\address{$^2$Department of Physics, University of California,
Santa Barbara, CA 93106-9530 USA}

\date{\today}

%\maketitle

\begin{abstract}

Experiments on polarized fermion gases performed by trapping ultracold
atoms in optical lattices, allow the study of an attractive Hubbard model for 
which the strength of the on site interaction is tuned by means of a Feshbach 
resonance. Using a well-known particle-hole transformation we discuss how
results obtained for this system can be reinterpreted in the 
context of a doped repulsive Hubbard model. In particular we show that
the Fulde-Ferrell-Larkin-Ovchinnikov (FFLO) state corresponds to the 
striped state of the 
two-dimensional doped positive U Hubbard model.
We then use the results of numerical studies of the striped state to relate the
periodicity of the FFLO state to the spin polarization. We also comment on 
the relationship
of the $d_{x^2-y^2}$
superconducting phase of the doped 2D repulsive Hubbard model to a d-wave 
spin density wave state for the attractive case.

\end{abstract}

  
\maketitle

Using standing wave laser light fields and optical Feshbach resonances, 
ultracold
atomic gases can be used to realize a variety of Hubbard like 
models.\cite{ref:1,ref:1a} Technically if one wants a strong interaction, 
it appears easier to create an
ultracold fermi gas (e.g., $^6$Li) with an attractive interaction and
experiments on polarized fermion gases with attractive interactions have
been carried out to look for exotic superfluid 
states.\cite{ref:2,ref:3,ref:4,ref:5}

One state of particular interest is the FFLO state which
arises from pairing across the spin-split Fermi surface of a spin polarized
system.\cite{ref:6,ref:7} In this state 
the Cooper pairs have a finite center of mass 
momentum leading to spatial oscillations
of the pair field order parameter and the spin polarization. Here, making 
use of density-matrix-renormalization-group
(DMRG) results\cite{ref:8,ref:9} for the doped, positive U Hubbard model and a
well-known particle-hole
transformation,\cite{ref:em,ref:hi,ref:sh,ref:ss} we discuss the properties 
of the FFLO state for the case of a half-filled
spin polarized negative U Hubbard model on a two-dimensional lattice. 
We also comment on
the relationship of this to the question of whether the doped positive U 
Hubbard model may have a $d_{x^2-y^2}$ superconducting ground state.

The Hamiltonian for a half-filled negative U Hubbard model in an external 
Zeeman field, $H$,  can be written as

\begin{equation}
  {\cal H}=-t\sum_{\langle ij\rangle s}(c^{\dagger}_{is}c_{js}+c^{\dagger}_{js}c_{is})+\frac{U}{2}\sum_i(n_{i\uparrow}-
  n_{i\downarrow})^2 
\nonumber 
\end{equation}
\begin{equation}
-h\sum_i(n_{i\uparrow}-n_{i\downarrow}).
  \label{eq:1}
\end{equation}

\noindent Here $t$ is a nearest neighbor one-electron hopping, $U$ is 
positive so 
the onsite interaction
is $-Un_{i\uparrow}n_{i\downarrow}$ and $h=g\mu_0H/2$. 
The operator $c^{\dagger}_{is}$
creates a fermion with spin $s$ on site $i$, the sum $\langle ij\rangle$ 
is over
nearest neighbor sites of a square lattice and 
$n_{is}=c^{\dagger}_{is}c_{is}$ is the number
operator.  In the case of the ultracold gases, 
a spin polarization $p=(N_\uparrow-N_\downarrow)/N$
is achieved by loading more atoms in the ``up'' pseudo-spin hyperfine state 
than in the ``down'' state. 

\begin{figure}[thbp]
\begin{center}
\includegraphics[width=8cm,clip,angle=0]{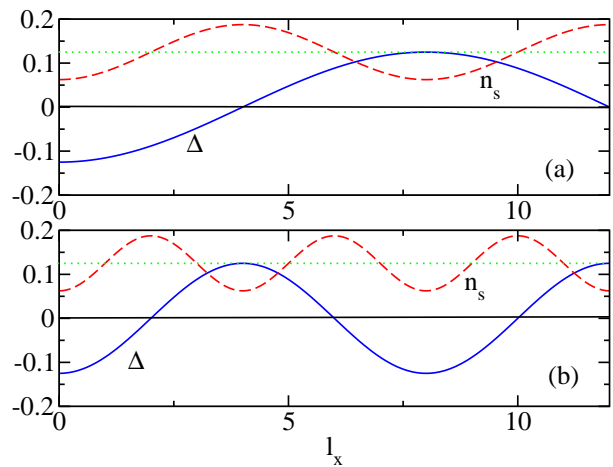}
\vskip 0.3cm
\caption{(Color on line) Schematic results for the spatial variation of 
the $s$-wave order
parameter $\Delta(\ell_x)$ (solid) and the spin polarization $n_s(\ell_x)$ 
(dashed)
in an FFLO state with $p=0.125$. (a) The mean field result with a period for
$\Delta(\ell_x)$  of $2/p=16$ and (b) with a period $1/p=8$ expected from
DMRG calculations on long stripes.}
\label{fig:1}
\end{center}
\end{figure}

In the mean-field FFLO state, the Cooper pairs have a 
finite center of mass momentum which, for a two dimensional tight binding 
band structure, lays along the (1,0) or the (1,1) direction depending upon 
the ratio of $|U|/t$. For the case in which it
lays along the (1,0) direction, the order
parameter exhibits a one-dimensional oscillation along the $x$ direction

\begin{equation}
\Delta(\ell_x)=Re\langle c_{\ell\uparrow}c_{\ell\downarrow}\rangle=
-\Delta_0\cos(q_x\ell_x).
  \label{eq:2}
\end{equation}
Here we have chosen the phase so that $\Delta(0)$ is negative.
The spin polarization also varies in space with

\begin{equation}
  n_s(\ell_x)=\langle n_{\ell\uparrow}-n_{\ell\downarrow}
\rangle=  p-m_0\cos(2q_x\ell_x).
  \label{eq:3}
\end{equation}
Here $\cos(q_x\ell_x)$ and $\cos(2q_x\ell_x)$ give the leading weak coupling 
harmonic behavior. 
The spatial variation of $\Delta(\ell_x)$ and $n_s(\ell_x)$ are
schematically illustrated in Fig.~\ref{fig:1}. At stronger coupling higher
harmonics can enter giving a more localized behavior.

Since the possibility of an FFLO state
was originally proposed,\cite{ref:6,ref:7} there has been great interest in 
determining whether it exists and
exploring its properties.  In particular, with the possibility of observing 
such a state in ultracold atomic gases, there have been a number of 
theoretical 
calculations.\cite{ref:cr,ref:pwy,ref:sr,ref:mmi,ref:sr2,ref:hjz,ref:hccl}
These calculations have been based upon a mean field description.
Here we are interested in the case of a two-dimensional half-filled lattice gas
where it is important to treat strong coupling effects. One would like to know
if the FFLO state survives
beyond the mean field approximation, whether the stripes run vertically 
(or horizontally) as in Eqs.~(\ref{eq:2})
and (\ref{eq:3}) or diagonally, and whether the periodicity of the spatial 
variation which determines $q_x$ remains at its mean field value.

The Hamiltonian for the negative U Hubbard model in a Zeeman field can be 
transformed under a unitary transformation to a doped positive U Hubbard model.
This transformation, introduced by Emery,\cite{ref:em}
has been a staple in quantum Monte Carlo work where it was used to show that 
the fermion sign problem is absent for the half-filled Hubbard model with a 
near-neighbor hopping.\cite{ref:hi} It has also
been used, as we will review, to provide a map between the half-filled 
negative U Hubbard model
in a Zeeman field and the doped positive U Hubbard 
model.\cite{ref:hi,ref:sh,ref:ss,ref:20}  Under this transformation
we will see that the FFLO state becomes the familiar striped, charge density, 
and $\pi$-phase shifted antiferromagnetic state of the doped positive U 
Hubbard model.

For a unitary transformation in which

\begin{equation}
  c_{\ell\uparrow}\to (-1)^{\ell_x+\ell_y}d^{\dagger}_{\ell\uparrow}
  \hspace{5mm}c_{\ell\downarrow}\to d_{\ell\downarrow}
  \label{eq:4}
\end{equation}

\noindent the Hamiltonian given in Eq.~(\ref{eq:1}) becomes

\begin{equation}
  {\cal H}=-t\sum_{\langle ij\rangle s}(d^{\dagger}_{is}d_{js}+d^{\dagger}_{js}d_{is})+\frac{U}{2}\sum(n_{i\uparrow}+
  n_{i\downarrow}-1)^2
\nonumber 
\end{equation}
\begin{equation}
-\mu\sum_i(n_{i\uparrow}+n_{i\downarrow})
  \label{eq:5}
\end{equation}

\noindent with $\mu=-h$. That is, the half-filled negative U Hubbard model with spin polarization
$p$ is mapped into a positive U Hubbard model with a site filling $\langle n\rangle=1-p$.
Under this transformation Eq.~(\ref{eq:2}) becomes

\begin{equation}
  (-1)^{\ell_x+\ell_y}\langle M_x(\ell_x,\ell_y)\rangle=-\Delta_0\cos q_x\ell_x
  \label{eq:6}
\end{equation}

\noindent with $M_x(\ell)=(d^{\dagger}_{\ell\uparrow}d_{\ell\downarrow}+d^
{\dagger}_{\ell\downarrow}d_{\ell\uparrow})/2$. The system has 
rotational symmetry so 
that Eq.~(\ref{eq:6}) implies that the staggered spin order is modulated by
$\cos q_x\ell_x$.

\begin{align}
   m_{stag}(\ell_x)&={(-1)^{\ell_x+\ell_y}\over{2}}\langle n_{\uparrow}
(\ell_x,\ell_y)-n_{\downarrow}(\ell_x,\ell_y)\rangle \nonumber\\
&=-\Delta_0\cos q_x\ell_x.
  \label{eq:7}
\end{align}

\noindent Similarly, the spin polarization, Eq.~(\ref{eq:3}), 
transforms to the hole 
density giving
\begin{equation}
   h(\ell_x)=1-\langle n(\ell_x,\ell_y)\rangle=p-m_0\cos(2q_x\ell_x).
  \label{eq:8}
\end{equation}
Thus the FFLO state maps to the striped state of the doped positive U Hubbard 
model in which charged stripes separate $\pi$-phase shifted antiferromagnetic 
regions.

Mean field studies\cite{ref:sh,ref:B,ref:A,ref:gi,ref:in,ref:C} of the doped,
two-dimensional positive U Hubbard
model with a nearest-neighbor hopping find that the domain-walls run along 
the $x$ or $y$-directions for
$|U|/t\ltwid3.6$.  For values of $3.6\ltwid|U|/t\ltwid8$ the domain-walls 
run diagonally and for $|U|/t$ greater than 8 the domain-walls are found to 
be unstable with respect to the
formation of magnetic polarons.\cite{ref:in2} The mean field domain walls are 
found to contain one hole per site along the wall and one says that the 
domain wall is filled with holes.  This means that the spacing of the vertical 
(or horizontal) charged domain walls is equal to $p^{-1}$, giving the
mean field result
\begin{equation}
  q_x=\pi p.
  \label{eq:9}
\end{equation}
for such walls. A schematic illustration of the mean field results for 
$p=0.125$ is shown in Fig.~\ref{fig:1}a.
 
Now these are mean field results and at larger values of U one is dealing 
with a strongly correlated system. Thus one might wonder whether the FFLO 
state remains the ground state. This is a question that has also
played a central role in the high $T_c$ cuprate problem.
There one would like to know if the ground state of the doped Hubbard model is
striped or possibly a $d_{x^2-y^2}$ superconductor, and what is the nature of
the interplay between stripes and pairing.\cite{ref:kiv} Present DMRG 
calculations on 6-leg
Hubbard ladders find a striped ground state.\cite{ref:8,ref:9}, while dynamic
cluster Monte Carlo calculations\cite{ref:13} on periodic lattices find evidence of a $d_{x^2-y^2}$
superconducting state. Whether the difference of lattice aspect ratios
and boundary conditions
or subtle numerical biases are responsible for this disagreement is not known.
It does, however, appear that these two states are very close in energy. Here
we will use the DMRG results for the striped state to discuss the nature of the
FFLO state and then conclude by noting what would happen for the 
$d_{x^2-y^2}$ state.

\begin{figure}[thbp]
\begin{center}
\includegraphics[width=8cm,clip,angle=0]{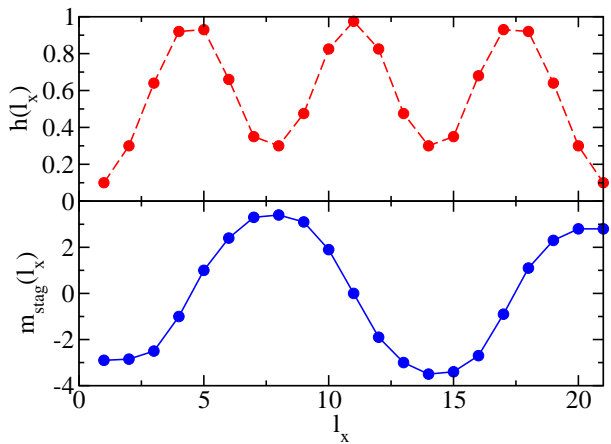}
\vskip 0.3cm
\caption{(Color online).
This figure shows DMRG results (G. Hager et al.\protect\cite{ref:9})
for the hole $h(\ell_x)$ (dashed red) and the staggered spin $m_{stag}(\ell_x)$
(solid blue) densities along the leg direction for a
$21\times6$ Hubbard ladder with 12 holes and $U/t=12$. As discussed
in the text, $h(\ell_x)$ corresponds to the spin polarization $n_s(\ell_x)$
%$\langle M_z(\ell_x)\rangle$
and $m_{stag}(\ell_x)$ corresponds to the $s$-wave pairfield order
parameter $\Delta(\ell_x)$ of the FFLO state.}
\label{fig:2}
\end{center}
\end{figure}

The DMRG calculations have been carried out on 6-leg doped positive
U Hubbard ladders. Periodic boundary conditions have been imposed in the 6-leg
direction and open boundary conditions are used in the long $x$-direction.  
In Ref.~10, the
results have been extrapolated to infinite length ladders. These numerical
results find that the ground state exhibits the striped structure shown
in Fig.~\ref{fig:2} for a $6\times21$ ladder with $U/t=12$ and 12 holes 
corresponding to a doping $p=2/21$. The figure shows the hole density
\begin{equation}
   h(\ell_x)=\frac{1}{6}\sum_{\ell_y=1}^6 (1-\langle n(\ell_x,\ell_y)\rangle),
  \label{eq:10}
\end{equation}
and the staggered spin density
\begin{equation}
   m_{\rm stag}(\ell_x)=\frac{1}{6}\sum_{\ell_y=1}^6 (-1)^{\ell_x+\ell_y}
\langle n_{\uparrow}
(\ell_x,\ell_y)-n_{\downarrow}(\ell_x,\ell_y)\rangle,
  \label{eq:11}
\end{equation}
 versus $\ell_x$.
For this doping, extrapolated DMRG results\cite{ref:9}
show that the ground state is striped for $U/t\gtwid 3$.
For a 6-leg tube, the stripes are found to contain 4 holes (2/3 filled stripes)
so that the charge stripe spacing is 
$2/{(3p)}=7$ with $\langle n\rangle=1-p$.

Calculations on an 8-leg $t-J$ ladder find that the stripes are half-filled 
and there is evidence that this is the preferred filling.\cite{ref:14} These
calculations show that the holes that make up the stripe exhibit short
range pairing correlations\cite{ref:30} in which the sign of the singlet
pair field associated with two sites connected by a hopping perpendicular to 
the stripe is opposite to that of the case in which the hopping is parallel 
to the stripe.
These d-wave like pairing correlations of the holes in the stripe are not taken into account 
in the mean
field calculations which find filled stripes. For half-filled stripes, the
spacing between the charged stripes is $1/2p$ so that the numerical 
calculations on the
longer stripes imply that $q_x$ for the FFLO state will be twice that found in
the mean field calculations.

\begin{equation}
  q_x=2\pi p.
  \label{eq:12}
\end{equation}

In this case, the FFLO state will have half the spatial period of the mean 
field solution, as illustration in Fig.~\ref{fig:1}b.

Here we have made use of DMRG calculations which were carried out on systems 
with
periodic boundary conditions in one direction and open boundary conditions 
in the other.
In cold atoms experiments, one is generally dealing with an underlying 
optical lattice
in a slowly varying trap potential which acts as a spatially dependent 
chemical potential.
This leads to soft boundaries and the simultaneous coexistence of spatially 
separated phases.  The details of this depend upon the total filling as well 
as the curvature
of the confining trap potential.\cite{ref:15} In principle, the DMRG 
method is well
suited to treating such soft boundaries but here we have focused our 
discussion on
the idealized case of a two-dimensional lattice which has one fermion per 
site with an
attractive Feshbach tuned interaction and a spin polarization $p$.  We have
shown that the FFLO state of a 2D half-filled attractive spin polarized 
Hubbard model is a unitary transform of the striped state of the doped 
repulsive Hubbard model. Thus an experimental observation of the FFLO state 
in the above mentioned cold atom set-up would imply that the 2D doped
repulsive Hubbard model is striped. 
Note that we have not shown that the striped
state is the ground state of the two-dimensional doped repulsive U Hubbard 
model. This remains an open question. We
have simply shown that if the striped state were the ground state then this 
would imply that
the FFLO state would be found in the optical lattice experiments and that 
its periodicity
would be given by Eq.~(\ref{eq:12}). 
We believe that the factor
of two increase in $q_x$ represents an important change from the mean field 
result
and reflects the tendency towards $d_{x^2-y^2}$ pairing.\cite{ref:30} 
Recent experiments also suggest that local pairing correlations are present 
on the cuprate stripes.\cite{ref:31}

Finding evidence for an FFLO state in a cold attractive spin polarized fermi lattice gas would provide evidence that the low temperature phase of the doped positive U Hubbard model is striped.
If on the other hand, the $d_{x^2-y^2}$ superconducting state describes the low
temperature phase of the repulsive Hubbard model, then the transformation,
Eq.~(\ref{eq:4}), tells us that the cold atom system will exhibit a
``d-spin density wave'' ground state characterized by an order parameter

\begin{equation}
\sum_k(cos k_x-cos k_y)\langle c^{\dagger}_{k+Q \uparrow}c_{k \downarrow}
\rangle
  \label{eq:13}
\end{equation}
\noindent with $Q=(\pi,\pi)$. Thus experiments on half-filled, spin polarized 
cold fermi
lattice gases with attractive on site interactions and the search for the FFLO
state have important implications for the properties of the doped repulsive U
Hubbard model. Likewise, our present understanding of the repulsive
Hubbard model can provide information relevant to experiments on the spin
polarized attractive fermi gas.

\section*{Acknowledgements}

We would like to thank T.~Esslinger, E.~Mueller, and M.~Holland for helpful 
discussions regarding the
recent experimental work on cold fermion gases and L.~Balents, S.A.~Kivelson,
and I.~Affleck for useful discussions regarding the FFLO state. We would also
like to thank G.~Hager, G.~Wellein, E.~Jeckelmann and H.~Fehske for providing 
the data for Figure 2.
A.M. is supported by NSF under grants DMR-0443144 and DMR-0454504,and partial 
support under Contract DE-AC05-00OR22725 with UT-Battelle, LLC.
D.J.S. would like to acknowledge the
Center for Nanophase Material Science at Oak Ridge National Laboratory for
support.


\begin{thebibliography}{99}


\bibitem{ref:1} D.~Jaksch {\it et al.}, Phys. Rev. Lett.{\bf 81}, 3108 (1998).

\bibitem{ref:1a} M. Holland {\it et al.}, Phys. Rev. Lett.{\bf 87}, 120406 
(2001).

\bibitem{ref:2} M.~K\"ohl {\it et al.}, Phys. Rev. Lett.{\bf 94}, 80403 (2005).

\bibitem{ref:3} M.W.~Zwierlein {\it et al.}, Science {\bf 311}, 492 (2006).

\bibitem{ref:4} G.B.~Partridge {\it et al.}, Science {\bf 311}, 503 (2006).

\bibitem{ref:5} Y.~Shin {\it et al.}, cond-mat/0606432.

\bibitem{ref:6} P.~Fulde and R.A.~Ferrell, Phys. Rev.{\bf 135}, A550 (1964).

\bibitem{ref:7} A.I.~Larkin and Y.N.~Ovchinnikov,  Sov. Phys. JETP {\bf 20}, 
762 (1965).

\bibitem{ref:8} S.R.~White and D.J.~Scalapino,  Phys. Rev. Lett. {\bf 91}, 
136403 (2003).

\bibitem{ref:9} G.~Hager {\it et al.}, Phys. Rev. {\bf B71}, 75108 (2005).

\bibitem{ref:em} V.J.~Emery, Phys. Rev. {\bf B14}, 2989 (1976).

\bibitem{ref:hi} J.E.~Hirsch, Phys. Rev. {\bf B31}, 4403 (1985).

\bibitem{ref:sh} H.J.~Schulz, J. Phys. France {\bf 50}, 2833 (1989).

\bibitem{ref:ss} M.I.~Salkola and J.R.~Schrieffer, Phys.Rev.{\bf B57},14433 (1998).

\bibitem{ref:cr} J.~Carlson and S.~Reddy, Phys. Rev. Lett. {\bf 95}, 
60401 (2005).

\bibitem{ref:pwy} C.-H.~Pao, S.-T.~Wu and S.-K.~Yip, Phys. Rev. {\bf B73}, 
132506 (2006).

\bibitem{ref:sr} D.E.~Sheehy and L.~Radzihovsky, Phys. Rev. Lett. {\bf 96}, 
60401 (2006).

\bibitem{ref:mmi} K.~Machida, T.~Mizushima and M.~Ichioka, Phys. Rev. Lett. 
{\bf 97}, 120407 (2006). 

\bibitem{ref:sr2} D.E.~Sheey and L.~Radzihovsky, cond-mat/0607803, 0608172.

\bibitem{ref:hjz} L.~He, M.~Jin and P.~Zhuang, cond-mat/0606322.

\bibitem{ref:hccl} Y.~He, C.-C.~Chien, Q.~Chen and K.~Levin, cond-mat/0610274.

\bibitem{ref:20} R. Scalettar {\it et al.},  Phys. Rev. Lett. {\bf 62}, 
1407 (1989).

\bibitem{ref:A} J.~Zaanen and O.~Gunnarsson, Phys. Rev.{\bf B40}, 7391 (1989).

\bibitem{ref:B} D.~Poilblanc and T.M.~Rice, Phys. Rev. {\bf B39}, 9749 (1989).

\bibitem{ref:gi} T.~Giamarchi and C.~Lhuillier, Phys. Rev. {\bf B42}, 10641 
(1990).

\bibitem{ref:in} M.~Inui and P.B.~Littlewood, Phys. Rev.{\bf B44}, 4415 (1991).

\bibitem{ref:C} Q.~Wang, H.-Y.~Chen, C.-R.~Hu and C.S.~Ting, Phys. Rev. 
Lett. {\bf 96}, 117006 (2006).

\bibitem{ref:in2} Here we have quoted results that Inui and 
Littlewood\protect\cite{ref:in} obtained from a Hartree-Fock study of a 
two-dimensional Hubbard model with a near neighbor hopping.

\bibitem{ref:kiv} S.A.Kivelson and E. Fradkin, cond-mat/0507459.

\bibitem{ref:13} T.A.~Maier {\it et al.}, Phys. Rev. Lett. {\bf 95}, 
237001 (2005).

\bibitem{ref:14} S.R.~White and D.J.~Scalapino,  Phys. Rev. Lett. {\bf 91}, 
3227 (1998).

\bibitem{ref:30} S.R.~White and D.J.~Scalapino, ``Why do stripes form in 
doped antiferromagnets and what is their relationship to superconductivity?", 
cond-mat/0006071. 

\bibitem{ref:15} M.~Rigol and A.~Muramatsu, Phys. Rev.{\bf A69}, 53612 (2004).

\bibitem{ref:31} T.~Valla {\it et al.}, {\it Sciencexpress Report}, 
16 November 2006, page 1.




\end{thebibliography}
\end{document}